# An audio CAPTCHA to distinguish humans from computers

Haichang Gao, Honggang Liu, Dan Yao, Xiyang Liu
Software Engineering Institute
Xidian University
Xi'an, Shaanxi 710071, P.R.China
hchgao@xidian.edu.cn

Uwe Aickelin
School of Computer Science
The University of Nottingham
Nottingham, NG8 1BB, U.K.
uxa@cs.nott.ac.uk

*Abstract*—CAPTCHAs are employed as a security measure to differentiate human users from bots. A new sound-based CAPTCHA is proposed in this paper, which exploits the gaps between human voice and synthetic voice rather than relays on the auditory perception of human. The user is required to read out a given sentence, which is selected randomly from a specified book. The generated audio file will be analyzed automatically to judge whether the user is a human or not. In this paper, the design of the new CAPTCHA, the analysis of the audio files, and the choice of the audio frame window function are described in detail. And also, some experiments are conducted to fix the critical threshold and the coefficients of three indicators to ensure the security. The proposed audio CAPTCHA is proved accessible to users. The user study has shown that the human success rate reaches approximately 97% and the pass rate of attack software using Microsoft SDK 5.1 is only 4%. The experiments also indicated that it could be solved by most human users in less than 14 seconds and the average time is only 7.8 seconds.

## I. INTRODUCTION

With the expansion of Internet, a great many daily activities are now done through Internet for convenience, including communication, education and e-commerce. As a matter of fact, web sites must ensure that the services are supplied to legitimate human users rather than bots to prevent service abuse. Most of them ask users to challenge puzzles before they are authenticated to the service. The puzzles, which are first introduced by Luis von Ahn et al. in 2003 [1], are CAPTCHAs.

CAPTCHA stands for *Completely Automated Public Turing Test to Tell Computers and Human Apart*, which is universally a secure measure to differentiate human users from bots and adopted by many websites. CAPTCHA is an automated Turing test that can generate and grade tests which human can easily pass while bots cannot. The existing CAPTCHAs can be generally classified into three categories: Text-based CAPTCHAs [4, 5], Image-based CAPTCHAs [8, 10] and Sound-based CAPTCHAs [12]. Text-based CAPTCHAs relay on the distortion of digits/letters and other visual effects added in the background image. The content can be a word or random alphanumeric characters. The user is asked to identify the distorted characters and entered them.

Image-based CAPTCHAs ask users to enter proper labels which can describe the image properly or require users to rotate the image to the correct direction. So the images and the corresponding labels which are not unique should be pre-stored. Sound-based CAPTCHAs are based on the auditory perception of human users, and can be divided into two categories. The first ones present users with a sound clip which contains distorted numbers and characters with background noise. The other kind offers sounds related with images. Current sound-based CAPTHAs have been broken by high-quality voice recognition and noise removal programs [2]. Some existing audio CAPTCHA is highly error prone and time consuming [3].

In contrast with the traditional sound based CAPTHCAs, our new proposed CAPTHCA exploits the gaps between human voice and synthetic voice. It generates challenges by presenting a sentence which is randomly selected from some books. The user is asked to read out the sentence, and then the mechanism estimates weather it is a human or not by analyzing the generated audio file. A user study was conducted to investigate the performance of our proposed mechanism. The result is encouraging.

## II. RELATED WORKS

It has been years for websites to use CAPTCHAs as a security measurement to distinguish human users from bots. So far, most commonly used CAPTCHAs are text-based CAPTCHAs. They are easily understood and solved because of their intuition. Also, they can be easily designed and implemented. Examples of these CAPTCHAs include EZ-Gimpy, which is deployed on Yahoo!; Pessimal Print [4], which contains common English words between 5-8 characters long; Baffletext [5], which uses non-English pronounceable strings and image-masking degradations. A big problem of these CAPTCHAs encountered is the development of automated computer vision techniques, which have been designed to remove noise and segment the distorted strings to make characters readable for OCR [6, 7]. As a matter of the fact, a number of text-based CAPTCHAs used previously on sites such as Yahoo! and Google have already been broken [2].

Image-based CAPTCHAs require users to identify labeled images or rotate images. It evinces a larger gap between human users and bots, because of the poor ability of bots in obtaining features of images. The early one was introduced as Bongo [8]. It puts forward a visual pattern recognition problem to users. Pix [8], all of whose images are pictures of concrete objects, ask the users to answer the question "what are these pictures of?" The difficulty with these CAPTCHAs is that they require a priori knowledge of the image labels [9]. Furthermore, all the images should be saved which may take up large storage space. An ensuing mechanism, Asirra [10], has taken it into consideration. Its image database is provided by website Petfinder.com, which is the largest website in world devoted to find home for homeless animals. What users need to do is to identify cats out from a set of pictures. Unfortunately, it was also proved to be broken at 82.7% rate in telling cats out from the given pictures [11].

Sound-based CAPTCHAs exploit a border range of human ability, which are mainly based on the auditory perception of human to identify words or letters in a sound clip after being distorted and adding background noise. Their emergence facilitates the use of vision impaired people. A typical sound-based CAPTCHA is reCAPTCHA proposed by the Carnegie Mellon University and later acquired by Google. Besides its text version, this mechanism also provides users with an audio clip in which eight numbers are spoken by different individuals. These sound based CAPTCHAs have been deployed on the google.com, dig.com and so on. A representative scheme of sounds related with images is Audio/image by Graig Sauer et al. [12], which combines audio sounds and visual images. However, solutions for test samples of current audio CAPTCHAs from popular Web sites have been achieved with accuracy up to 71% [13]. It also proved that existing audio CAPTCHAs are more difficult and time-consuming [3].

The above-mentioned CAPTCHA schemes have not provided a satisfactory answer for usability and security, the two of major design and implementation issues of CAPTCHAs. In this paper, a new audio CAPTCHA which exploits the gaps between human voice and synthetic voice rather than relays on the auditory perception of human is proposed to solve the previous questions.

### III. THE PROPOSED SCHEME

The new CAPTCHA was implemented using C# and SilverLight. Figure 1 is the interface when the program is running. The waveform diagram can display the wavelet of the voice and give a feedback to the user what they have said. The sentence displayed in the textbox is the content to be read out. All sentences are generated randomly from some books, including History, Biography and others. In our scheme, the lengths of the sentences are set between 8 and 20 words.

After the entering of voice, the scheme will judge whether the user is a human or spyware by analyzing the temporarily stored audio file. In analysis step, there are some specified indicators, which will be described in detail in Section 4, adopted to classify the audio file. If the authentication ends with success, the user will be allowed to access the specific information.

There are some significant contributions for our scheme. First, it's the first time to exploit the gaps between human voice and synthetic voice as the base of a CAPTCHA. The previous audio CAPTCHAs are generally composed of a set of words to be identified, layered on top of noise. It is more scientific to judge a user is human or not by analyzing the audio characteristics.

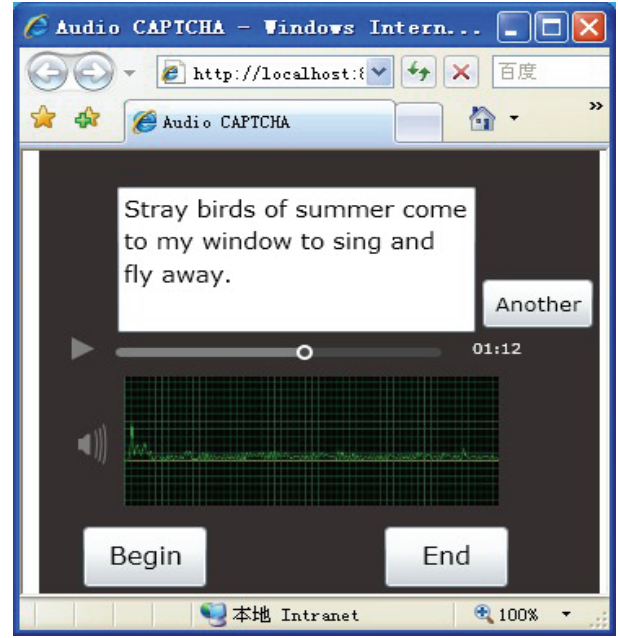

Figure 1. The interface of proposed audio CAPTCHA.

Second, our CAPTCHA provides a quite satisfactory security. The security is often associated with the challenges space of a scheme. The traditionally audio CAPTHCA schemes often store many audio files in advance, which make the challenges limited. However, the challenge in our scheme is generated from text files and needs much less space, and the space can be enlarged by expanding the text files library.

Third, the usability is encouraging. Result shows that the success rate of natural voice reaches approximately 97%, and the pass rate of synthesized voice which using Microsoft SDK 5.1 is only 4%.

### IV. THE ANALYSIS OF AUDIO FILE

In our scheme, short-term Fourier analysis is used to extract the characteristics of the audio file. Short-term Fourier analysis divide speech stream into segments for processing, each segment is named a "frame". Frame size ranges from 10ms to 30ms, and usually is 20ms. In order to reduce the truncation effect of speech frames, window-adding treatment is needed.

There are many conventional window functions. Normally, a tapered window function, such as Rectangular, Hanning or Hamming, has been used in earlier studies [16].

## A. The choice of the window function

Window functions can truncated the audio frame and make it smooth. Appropriate window function can raise the performance of the FFT (fast Fourier transform) under the same order. It means that appropriate window functions can reduce the number of filter bands in meeting the requirements of the situation. The standards of choosing window include:

(a) Low side lobe amplitude, especially the first side lobe;
(b) Side lobe amplitude should decline more rapidly in order to increase the Stop-band attenuation.

All the requirements can not be met at the same time, and actual is often being the compromised one. Hamming window is used the most popularly in the FFT technique. Hamming window has wider main lobe and smaller side lobes in comparison to rectangular window; the Hamming window provides better trade-off between frequency resolution and spectral leakage [17].

In our scheme, the signal is processed by adding the hamming window with the window width of 50 data. The formula used is shown as follow [14, 15]:

$$S(i) = w(n) * x(i) \quad (1)$$

$$w(n) = (0.54 - 0.46 \times \cos((2 \times PI \times n)/(N-1))),$$
$$0 \leq n \leq N-1 \quad (2)$$

$N$ is the length of the window, also the window data. $N$ is 50 in our scheme. $n$ is the location of the sample in the window. $x(i)$ is the input signal, $S(i)$ is the output signal after the treatment of adding Hamming windows. After the initial signal is processed, data of some indicators will be extracted from the final signal.

## B. The Indicators

The indicators used in this scheme include Short-term energy, Short-term average amplitude and Short-term zero-crossing rate.

*1) Short-term energy:* Short-term energy of speech signal is a signal strength measurement parameter. Short-term energy can do voiced-voiceless distinction, since short-term energy of the voiced is much bigger than the voiceless. It also determines the boundary of consonants and vowels and the boundary of silence and sound. The formula used to calculate the energy is shown as below [18]:

$$E_0 = \sum_{n=0}^{N-1} S^2(i) \quad (3)$$

$n$ is from 0 to $N$-1, each stands for one sample of the audio. If the start point is not from beginning but $m$, then $n$ starts from $m$.

*2) Short-term average amplitude:* The audio data is actually the changes of amplitude by the Time-domain. But a major problem for short-tem energy function is that it is sensitive to the signal level value. As the need to calculate the square of the sample value, it is very easy to overflow in the fixed-point. In order to overcome this drawback, average amplitude is utilized to measure the magnitude of change in language. Short-term average amplitude is shown as following [18]:

$$M_0 = \sum_{n=0}^{N-1} |S_w(n)| \quad (4)$$

*3) Short-term zero-crossing rate:* Zero-crossing refers to the signal value if it is higher than zero or not. Zero-crossing rate is the total counts of the signals whose value is higher than zero in a second. For the discrete-time sequence, zero-crossing means sign changes of the samples, and zero-crossing rates is the counts of sign changes of each sample. As for speech signal, it refers to the times speech signal waveform across the horizontal axis in one frame. Short-term zero-crossing rate is shown as following [18]:

$$Z_0 = \frac{1}{2} \sum_{n=0}^{N-1} |Sgn(S_w(n)) - Sgn(S_w(n-1))| \quad (5)$$

At last, normalize the $E_0$, $M_0$, and $Z_0$ to be $E$, $M$ and $Z$. And the calculation of the target value relies on the three indicators by Formula 6:

$$V = a*E + b*M + c*Z \quad (6)$$

$a$, $b$, $c$ are the weight of the three parameters $E$, $M$ and $Z$, and the sum of them is 1. For each audio file, the target value $V$ can be calculated by the above formula. And the threshold $V'$ will be definite in experiments. If $V$ is higher than $V'$, the user will be thought as bots, otherwise, human.

## C. The processes of the analysis

The process of audio file in our scheme is following. First, divide the audio stream into sample frame by short-time Fourier analysis. The frame size is 20ms in this scheme, and the frame shift is half of the frame size.

Second, add window function. In order to reduce the truncation effect of speech frames, Hamming windows have been used to treat the frame. The length of the window sequence plays an important role in the effect of the treatment. We have chosen the size of the window is half of the frame size. Each window includes 50 samples.

Third, calculate the values of Short-term energy, Short-term average amplitude, and Short-term zero-crossing rate by the above formulas.

At last, analyze the audio file data by normalized analysis and calculate the final value V' using the parameters. Then, compare the value with the threshold to judge whether the user is human or bots.

## V. PARAMETERS DETERMINATION

Ten students from computer science major are invited to participant in the experiments to determine the parameters in formula (6), each one attempts to authentication for ten times. All the participants are university students, of which six are female. The average age of the participants was 25 years, ranged from 23 to 27.

This experiment displayed a random sentence each time. Users are asked to read the specific sentence out to be authenticated. The users' respond time and accuracy rate were recorded. Besides, the speech synthesis software using Microsoft Speech SDK 5.1 is adopted as the attackers.

All 100 samples login by human and audio software have been collected. During Section 4, the indicators' concepts of Short-term energy ($E$), Short-term average amplitude ($M$),

Short-term zero-crossing rate (*Z*) have been introduced. Table 1, 2 and 3 show the result decided by single indicator. The value of the first column in each table is the corresponding threshold for single indicator. The value of the second column is the misjudgment rate of natural speech which stands for the rate to misjudge the human user as bots. The last is the synthesized speech recognition rate which is the rate to identify the current user as bots correctly.

We think that the misjudgment rate of natural speech is lower than 10% and the recognition rate of synthesized speech is higher than 90% at the same time is accepted. The above tables shows that single indicator can not meet the requirement. Therefore, all three indicators should be considered at the same time in order to make the scheme more idealized. First, normalize all the three kinds of data. For each audio file, we define *V* as the target value calculated from the Formula 6.

TABLE I. RESULTS DECIDED BY SHORT-TERM ENERGY

| Value | The misjudgment rate of natural speech | The recognition rate of synthesized speech |
|---|---|---|
| 4.6E+14 | 19% | 100% |
| 5.5 E+14 | 15% | 90% |
| 6.0E+14 | 6% | 86% |
| 7.0E+14 | 4% | 80% |
| 8.0E+14 | 3% | 79% |
| 9.0E+14 | 3% | 75% |
| 1.02E+15 | 0% | 65% |

TABLE II. RESULTS DECIDED BY SHORT-TERM AVERAGE AMPLITUDE

| Value | The misjudgment rate of natural speech | The recognition rate of synthesized speech |
|---|---|---|
| 79903 | 94% | 100% |
| 1000000 | 73% | 69% |
| 1400000 | 54% | 51% |
| 2008000 | 42% | 45% |
| 4000000 | 19% | 32% |
| 7049500 | 12% | 27% |
| 14445326 | 0% | 15% |

TABLE III. RESULTS DECIDED BY SHORT-TERM ZERO-CROSSING RATE

| Value | The misjudgment rate of natural speech | The recognition rate of synthesized speech |
|---|---|---|
| 4.1E+22 | 11% | 100% |
| 1.0 E+23 | 10% | 95% |
| 2.0E+23 | 5% | 80% |
| 2.64092E+23 | 1% | 66% |
| 5.62423E+23 | 0% | 43% |

The parameters *a*, *b* and *c* are the weight values for corresponding indicators. The accuracy of authentication is affected by the values of *a*, *b* and *c*. And then, classified the file into natural voice or synthesized voice by the value of *V'*. For each set of *a*, *b* and *c*, the value of the *V'* is ranged from 0 to 1, Figure 2 gives one example, in which Y-axis represents the rate and X-axis represents the values of *V'*.

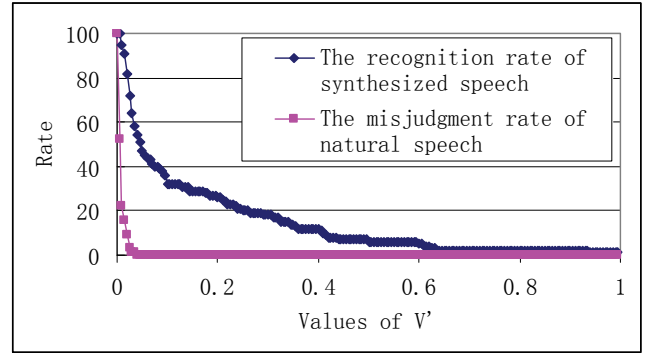

Figure 2. The recognition rate of synthesized speech and the misjudgment rate of natural speech.

In figure 2, *a* is set as 0.01, *b* is set as 0.88 and *c* is set as 0.11. The abscissa is the value of *V'*. The red curve stands for the misjudgment rate of natural speech and the blue stands for the synthesized speech recognition rate. If the synthesized speech recognition rate is high and the misjudgment rate of natural speech is low, the result is accepted. There are thousands of combination of *a*, *b*, *c* and *V'*. Table 4 shows some different combinations of the four parameters along with the synthesized speech recognition rate.

TABLE IV. RESULT FOR DIFFERENT COMBINATIONS OF *A*, *B*, *C* AND *V'*

| a | b | c | V' | The misjudgment rate of natural speech | The recognition rate of synthesized speech |
|---|---|---|---|---|---|
| 0 | 0.96 | 0.02 | 0.015 | 2% | 90% |
| 0.01 | 0.88 | 0.11 | 0.015 | 21% | 91% |
| 0.01 | 0.98 | 0.01 | 0.01 | 3% | 97% |
| 0.02 | 0.81 | 0.17 | 0.01 | 34% | 96% |
| 0.03 | 0.28 | 0.69 | 0.03 | 79% | 91% |
| 0.18 | 0.77 | 0.05 | 0.01 | 14% | 95% |
| 0.22 | 0.74 | 0.04 | 0.01 | 5% | 95% |

Finally, we chose *a* as 0.01, *b* as 0.98, *c* as 0.01 and *V'* as 0.01 by the experiments. The rate that human is refused is 3%, and the rate that the bots is identified is 97%.

After the parameters and the threshold were defined, 50 students were invited to use the scheme for 10 times, the result shows that the misjudgment rate of natural speech is 4%. Meanwhile, the Microsoft synthesized speech is used to check the security of the scheme. Totally, 100 synthesized speeches are analyzed and the recognition rate is 96%. Both the misjudgment rate of natural speech and the synthesized speech recognition rate are reasonable, which support that the values of the parameters and the threshold available.

## VI. USER STUDY

This user study was conducted to analyze the usability of our CAPTCHA. About 100 people were invited to participate in the study. The majority of the participants are college students aged from 20 to 25.

For the purpose of familiarizing the participants with this CAPTCHA, some guidance on how to operate was given. Then the participants were instructed to pass the challenges once. The time to complete one entire challenge was recorded. Simultaneously, a collection of another 50 sentences read by Microsoft Speech SDK 5.1 was conducted. The result shows that the success rate of natural voice reaches approximately 97%, and the success rate of synthesized voice is only 4%.

Table 5 indicates the time users spent to complete a challenge. It turned out that most people can finish it in 14 seconds, and all the participants completed the challenge within 18 seconds. The average completion time is 7.8 seconds. There are some reasons that make the audio CAPTCHA is time consuming than text-based CAPTCHA. First, the text-based CAPTHCA is usually 6 to 8 characters and user can enter them quickly. Second, when user read one sentence, he or she must start reading from the beginning again when one word is reading wrongly. However, our CAPTCHA is less time consuming compared to the traditional Sound-based CAPTCHAs.

After that, the participants were required to fill in the questionnaires like '*Which CAPTCHA scheme do you prefer?*' Results show the proposed CAPTCHA is relatively enjoyable. More than half (68%) participants enjoyed to complete this CAPTCHA than text-based CAPTCHA.

TABLE V. THE USERS' COMPLETION TIME

| Completion time | Users | Average completion time |
|---|---|---|
| <=6 sec | 20% | 7.8 sec |
| <=10 sec | 68% | |
| <=14sec | 89% | |
| <=18sec | 100% | |

## VII. CONCLUSIONS

In this paper, a novel sound based CAPTCHA is proposed. Unlike currently existed sound based CAPTCHAs, our CAPTCHA exploits the gaps between human voice and synthetic voice. The user only needs to read out a given sentence to pass the challenge. A user study has also been conducted to verify the usability of the CAPTHA. It has been proved that the success rate for human voice is approximately 97% and the attack success rate is 4%.

In this paper, only Microsoft synthesized speech is considered, but which kind of synthetic voice software is used in the actual circumstance is uncertain. Experiments were also carried on some other speech synthesis softwares, such as TextAloud and DSpeech. Under the same critical value (*a* as 0.01, *b* as 0.98, *c* as 0.01 and *V'* as 0.01), the results shows that the attack pass rate of TextAloud is 48%, and the rate of DSpeech is 39% which is not ideal. But the pass rate can be lower by changing the value of *a*, *b*, *c* or *V'* to some extent.

Future work will concentrate on finding a general method to distinguish kinds of synthetic speech softwares and the natural voice. Analysis and experiments are needed to improve the security of our scheme.

ACKNOWLEDGMENT

The authors would like to thank the reviewers for their constructive comments. Project 72105274 supported by the Fundamental Research Funds for the Central Universities.